%% file: foregroundcc.tex
\pdfoutput=1
\documentclass[amssymb,amsmath,prd,nofootinbib,twocolumn,superscriptaddress]{revtex4-1}
\usepackage{fontawesome5}
\usepackage{lipsum}
\usepackage{appendix}
\usepackage{diagbox}
\usepackage{booktabs}
\usepackage{physics}
\usepackage{amsmath}
\usepackage{amssymb}
\usepackage{mathrsfs}
\usepackage{tcolorbox}
\usepackage[nolist,nohyperlinks]{acronym}
\usepackage[caption=false]{subfig}
\usepackage{url}
\usepackage{multirow}
\usepackage{booktabs}
\usepackage{xcolor}
\definecolor{linkcolor}{HTML}{4A70A9}
\usepackage[%
  colorlinks = true,
  urlcolor= linkcolor,
  linkcolor= linkcolor,
  citecolor= linkcolor
]{hyperref}
\usepackage{orcidlink}
\usepackage{float}
\usepackage{graphicx}
\usepackage{capt-of}
\usepackage{tikz-cd}
\usepackage{tikz}
\usetikzlibrary{shapes,arrows}
\usetikzlibrary{shapes.geometric, arrows}
\usetikzlibrary{graphs}
\usetikzlibrary{cd}
\usepackage{dcolumn}
\usepackage{bm}

\newcommand{\D}{\mathrm{d}}
\newcommand{\seg}{\mathrm{seg}}
\newcommand{\imag}{\mathfrak{I}}
\newcommand{\Cdiag}{\hat C_f^{(I);\,\mathrm{diag}}}
\newcommand{\Coffdiag}{\hat C_f^{(I);\,\mathrm{off-diag}}}
\newcommand{\obs}{\mathrm{obs}}

\newcommand{\tot}{\mathrm{tot}}

\allowdisplaybreaks

\begin{document}
\title{All You Need is not \texorpdfstring{$\Omega_\mathrm{gw}$}{Omega\_gw}: Beyond the Mean of the Cross-Correlation Estimator when Searching for an Astrophysical Gravitational-Wave Background}

\author{Haowen Zhong\texorpdfstring{\,}{}\orcidlink{0000-0001-8324-5158}}
\email{zhong461@umn.edu}
\affiliation{School of Physics and Astronomy, University of Minnesota, Minneapolis, MN 55455, USA}

\author{Francesco Iacovelli\texorpdfstring{\,}{}\orcidlink{0000-0002-4875-5862}}
\email{fiacovelli@jhu.edu}
\affiliation{William H. Miller III Department of Physics and Astronomy, Johns Hopkins University, Baltimore, Maryland 21218, USA}

\author{Vuk Mandic\texorpdfstring{\,}{}\orcidlink{0000-0001-6333-8621}}
\email{vuk@umn.edu}
\affiliation{School of Physics and Astronomy, University of Minnesota, Minneapolis, MN 55455, USA}

\author{Emanuele Berti\texorpdfstring{\,}{}\orcidlink{0000-0003-0751-5130}}
\email{berti@jhu.edu}
\affiliation{William H. Miller III Department of Physics and Astronomy, Johns Hopkins University, Baltimore, Maryland 21218, USA}
\date{\today}
\begin{abstract}
Searches for the stochastic gravitational-wave background (SGWB) using ground-based detectors rely heavily on the cross-correlation estimator $\bar C_f$, whose statistical properties determine the theoretical foundation for inference frameworks built on it. Past analyses usually assume Gaussianity of $\bar C_f$, but this assumption has not been systematically tested for a background of astrophysical origin like the one produced by compact binary coalescences. In this work, we decompose $\bar C_f$ into three physically motivated components: (1) the mean intensity, (2) the geometrical shot noise, and (3) the polarization leakage, and we compute 90\% credible intervals for its cumulants up to fourth order for the binary black hole, binary neutron star, and neutron star-black hole populations by propagating local merger rate uncertainties. We find that for the upcoming LIGO O5 sensitivity, $\bar C_f$ remains effectively Gaussian for all three populations, validating current Gaussian likelihood frameworks. For Cosmic Explorer (CE), however, the $\bar C_f$ produced by binary black holes becomes non-Gaussian, with skewness and excess kurtosis of $0.31^{+0.04}_{-0.03}$ and $1.3^{+0.4}_{-0.3}$, respectively, in the most sensitive band over a one-year observation period. For binary neutron star and neutron star - black hole mergers, $\bar C_f$ remains largely Gaussian even at CE. These results indicate that the Edgeworth expansion provides an adequate leading-order description of the non-Gaussianity expected in the next-generation era, while also motivating the development of more sophisticated statistical methods for a more accurate treatment.
\end{abstract}
\maketitle
\begin{acronym}
    \acro{GW}{gravitational-wave}
    \acro{ORF}{overlap reduction function}
    \acro{PSD}{power spectral density}
    \acro{GR}{general relativity}
    \acro{CBC}{compact binary coalescence}
    \acro{BH}{black hole}
    \acro{BBH}{binary black hole}
    \acro{BNS}{binary neutron star}
    \acro{NSBH}{neutron star-black hole}
    \acro{LVK}{LIGO-Virgo-KAGRA}
    \acro{PE}{parameter estimation}
    \acro{SGWB}{stochastic gravitational wave background}
    \acro{CGWB}{cosmological gravitational wave background}
    \acro{CDF}{cumulative distribution function}
    \acro{CGF}{cumulant generating function}
    \acro{MGF}{moment generating function}
    \acro{MC}{Monte Carlo}
    \acro{CE}{Cosmic Explorer}
    \acro{XG}{next-generation}
    \acro{SNR}{signal-to-noise ratio}
    \acro{CLT}{central limit theorem}
    \acro{HDI}{highest density interval}
    \acro{PN}{post-Newtonian}
    \acro{PTA}{pulsar timing array}
\end{acronym}
\input{intro}
\input{cc_new}
\input{simulations}
\input{discussion}

\begin{acknowledgments}
The authors are grateful for computational resources provided by the LIGO Laboratory and supported by National Science Foundation (NSF) Grants PHY-0757058 and PHY-0823459. The authors acknowledge the Minnesota Supercomputing Institute (MSI) at the University of Minnesota for providing resources that contributed to the research results reported within this paper. URL: \url{http://www.msi.umn.eduledgments}. 
The work of H.Z. and V.M. was supported by the NSF grant PHY-2409173.
F.I. and E.B. are supported by NSF Grants No.~AST-2307146, No.~PHY-2513337, No.~PHY-090003, and No.~PHY-20043, by NASA Grant No.~21-ATP21-0010, by John Templeton Foundation Grant No.~62840, by the Simons Foundation [MPS-SIP-00001698, E.B.], by the Simons Foundation International [SFI-MPS-BH-00012593-02], and by Italian Ministry of Foreign Affairs and International Cooperation Grant No.~PGR01167.
This work was carried out at the Advanced Research Computing at Hopkins (ARCH) core facility (\url{https://www.arch.jhu.edu/}), which is supported by the NSF Grant No. OAC-1920103. 
The work of F.I. is supported by a Miller Postdoctoral Fellowship.

\end{acknowledgments}
\bibliography{references}
\end{document}

%% file: intro.tex
\section{Introduction}\label{sec:intro}
The direct detection of gravitational waves (GWs) by the LIGO–Virgo–KAGRA (LVK) network has opened a new observational window onto the Universe~\cite{LIGOScientific:2016aoc, LIGOScientific:2018mvr,LIGOScientific:2020ibl, KAGRA:2021vkt, LIGOScientific:2025slb, LIGOScientific:2026wfs}. Beyond resolved \acp{CBC}, an as-yet undetected \ac{SGWB} — the incoherent superposition of unresolved astrophysical and cosmological sources — encodes integrated information about the GW emission history of the Universe~\cite{Christensen:2018iqi, Renzini:2022alw, Romano:2016dpx}. The astrophysical SGWB sourced by unresolved CBCs is expected to dominate in the LVK band and is projected to be detected within the next decade~\cite{LIGOScientific:2025bgj, KAGRA:2021kbb}.

The standard approach to searching for the \ac{SGWB} using ground-based detectors is the cross-correlation method~\cite{Allen:1997ad, Romano:2016dpx}, in which the time-domain data from a detector pair is divided into segments and combined into the frequency-domain estimator $\bar C_f$. Most of the existing analyses, including the past LVK searches~\cite{KAGRA:2021kbb,LIGOScientific:2025bgj}, assume that $\bar C_f$ follows a Gaussian distribution; the likelihood then factorizes as a Gaussian in each frequency bin, and inference on the \ac{SGWB} amplitude reduces to standard matched-filter statistics~\cite{Mandic:2012pj, Matas:2020roi}. This Gaussian assumption is well motivated for a truly stochastic background composed of a \emph{large number} of incoherently superposed weak sources, and is therefore a good assumption for a cosmological \ac{SGWB}. 

For an astrophysical \ac{SGWB} sourced by \acp{CBC}, however, the picture is more subtle. The signal is composed of a finite number of discrete, transient events with intrinsic chirp-like time–frequency structure, definite polarization, and source-specific sky locations and intrinsic parameters. Whether the resulting $\bar C_f$ is Gaussian depends on how many such events contribute per analysis segment and how detector responses average over them. Several previous works have addressed aspects of this question: the popcorn / non-stationary regime has been studied analytically in Refs.~\cite{Drasco:2002yd, Smith:2017vfk,Biscoveanu:2020gds, Lawrence:2023buo, Kou:2025bhk, Mukherjee:2019oma,Sah:2023bgr,Sah:2025agw}, weak departures from Gaussianity have been incorporated via the Edgeworth expansion in Refs.~\cite{Martellini:2014xia, Martellini:2015mfr}, and the deviation of the detector response to individual events from the sky-averaged \ac{ORF} has been studied in Ref.~\cite{Liu:2026xhc}. Nevertheless, a systematic, quantitative characterization of the higher-order statistics of $\bar C_f$ — across the full frequency band, the relevant CBC populations, and different detector sensitivities — has so far been missing.

In this paper, we provide such a characterization. We decompose $\bar C_f$ into three physically distinct contributions: (1) the mean intensity, (2) the geometrical shot noise, and (3) the polarization leakage. We compute the cumulants $\kappa_n(\bar C_f)$ up to fourth order for the \ac{BBH}, \ac{BNS}, and \ac{NSBH} populations. We find that $\bar C_f$ remains effectively Gaussian for all three populations for the LIGO sensitivity in the upcoming fifth observation run (O5), justifying the continued use of Gaussian-likelihood frameworks for the foreseeable future. For \ac{XG} GW detectors, Einstein Telescope (ET)~\cite{Punturo:2010zz,Branchesi:2023mws,ET:2025xjr} and \ac{CE}~\cite{Reitze:2019iox,Evans:2021gyd,Evans:2023euw}, the situation changes qualitatively. For a network comprising two \ac{CE} instruments, the $\bar C_f$ from the \ac{BBH} background develops appreciable non-Gaussianity, with the skewness and excess kurtosis of $\bar C_f$ reaching $\mathcal{O}(0.3)$ and $\mathcal{O}(1.3)$, respectively, in the most sensitive band of the detector network. While the Edgeworth expansion --- the standard tool for incorporating mild non-Gaussianity --- remains adequate as a leading-order description in this regime, its limited range of validity motivates a more accurate treatment (or, alternatively, a notching strategy that removes resolved \ac{BBH} signals from the data stream~\cite{Zhong:2022ylh, Zhong:2024dss, Zhong:2025gxk}).

The remainder of this paper is organized as follows. Section~\ref{sec:cc} reviews the cross-correlation estimator and derives the cumulant expansion of $\bar C_f$ for a \ac{CBC}-sourced \ac{SGWB}.  Section~\ref{sec:sim} presents the numerical results for the \ac{BBH}, \ac{BNS}, and \ac{NSBH} backgrounds across the LIGO O5 and CE detector configurations. Section~\ref{sec:discussion_conclusion} discusses the implications of our findings, the breakdown of the diagonal approximation for BNSs at low frequencies, and frequency correlations as an additional caveat for future work.

%% file: cc_new.tex
\section{Cross-Correlation}\label{sec:cc}
This section outlines the cross-correlation search for a \ac{SGWB}. The first part assumes the target \ac{SGWB} is of cosmological origin, with the standard theoretical properties expected in the literature, i.e., persistent, Gaussian, stationary, isotropic, and unpolarized. In the second half, we consider a \ac{SGWB} of astrophysical origin, such as the one generated by \ac{CBC} mergers.
\subsection{Cross-correlation search for a SGWB of cosmological origin}
A \ac{CGWB} results from the \textit{incoherent} superposition of a large number of \ac{GW} sources originating in the early Universe~\cite{Caprini:2018mtu}. Because of its stochastic nature, this background must be described statistically. Conventionally, a \ac{CGWB} is characterized by its dimensionless energy density spectrum
\begin{equation}
    \Omega_{\mathrm{gw};f}\equiv \frac{f}{\rho_{c,0}}\frac{\D \rho_{\mathrm{GW};f}}{\D f}\;,
\end{equation}
where $\D\rho_{\mathrm{GW};f}$ is the \ac{GW} energy density in the frequency bin $(f,f+\D f)$, $\rho_{c,0}=3H_0^2c^2/(8\pi G)$ is the critical energy density to close the Universe, $c$ is the speed of light, $G$ is the gravitational constant, and $H_0$ is the Hubble constant. Through the paper, we assume a flat $\Lambda$CDM cosmological model with \textsc{Planck18} parameters~\cite{Planck:2018vyg}.

Because a \ac{CGWB} signal is statistically indistinguishable from intrinsic instrumental noise, extracting it with a single detector is impossible unless the detector noise is negligible compared to the signal or the noise is well understood. Instead, one must cross-correlate strain data across a network of detectors. Provided the noise sources are uncorrelated between detectors, the common \ac{CGWB} signal can then be filtered out. For simplicity, we consider a two-detector network. Letting $\tilde d_{a;f}$ denote the frequency-domain strain for the detector $a\in\{1,\,2\}$, we can write
\begin{equation}
\tilde d_{a;f}=\tilde h_{a;f}+\tilde n_{a;f}\;.
\label{eq:d_h_n}
\end{equation}
Here, $\tilde h_{a;f}$ and $\tilde n_{a;f}$ represent the frequency-domain signal and noise strains, respectively. One usually models the \ac{CGWB} signal as a persistent, Gaussian, isotropic, unpolarized, and stationary random process, while the detector noise is assumed to be Gaussian and uncorrelated across different detectors. Under these assumptions, the two-point correlators are given by
\begin{equation}
\begin{aligned}
 &\langle \tilde h_{a;f}^*\tilde h_{
    b;f'}\rangle=\frac{1}{2}\delta(f-f')\gamma_{ab;f}P_{\mathrm{gw};f}\;,\\
&\langle \tilde n_{a;f}^*\tilde n_{
    b;f'}\rangle=\frac{1}{2}\delta(f-f')\delta_{ab}P_{n_a;f}\;,\\
&\langle \tilde h_{a;f}^*\tilde n_{
    b;f'}\rangle=0\;,\\
\end{aligned}
\label{eq:correlator_continuous}
\end{equation}
where $\gamma_{ab;f}$ denotes the \ac{ORF} of two detectors~\cite{Allen:1997ad,Romano:2016dpx}, and it is normalized such that $\gamma_{ab;f}\equiv 1$ for two colocated and aligned detectors. 
$P_{n_a;f}$ represents the detector noise \ac{PSD}, and $P_{\mathrm{gw};f}$ is the strain \ac{PSD} of the \ac{CGWB}~\cite{Mingarelli:2019mvk}. $P_{\mathrm{gw};f}$ and $\Omega_{\mathrm{gw};f}$ are connected via the following relation:
\begin{equation}
\Omega_{\mathrm{gw};f}=\frac{10\pi^2f^3}{3H_0^2}P_{\mathrm{gw};f}\;.
\end{equation}
We note that Eq.~\eqref{eq:correlator_continuous} is written in terms of the continuous Fourier transform of $d_t,\,h_t,$ and $n_t$. However, since we deal with discrete data streams in real searches, we can only perform discrete Fourier transforms in practice. 
The definition of the discrete Fourier transform of an arbitrary time series $x_j$ we use is
\[
\tilde{x}_\ell\equiv \sum_{j=0}^{N_d-1}x_j\, e^{-2\pi i f_\ell t_j}\Delta t\;,\qquad
x_j=\sum_{\ell=0}^{N_d-1}\tilde{x}_\ell\, e^{2\pi i f_\ell t_j}\Delta f\;,
\]
where $N_d$ denotes the number of data points of the series, $\Delta t$ the time resolution, $\Delta f$ the frequency resolution, $f_\ell\equiv \ell\Delta f$, $\Delta f\equiv 1/\tau\equiv 1/(N_d\Delta t)$, and $\ell=0,1,\dots,N_d-1$.
The discrete form of Eq.~\eqref{eq:correlator_continuous} is given by~\cite{Zhong:2026owg}
\begin{equation}
\begin{aligned}
&\langle \tilde h_{a;f}^*\tilde h_{
    b;f'}\rangle\simeq\frac{\tau}{2}\delta_{ff'}\gamma_{ab;f}P_{\mathrm{gw};f}\;,\\
&\langle \tilde n_{a;f}^*\tilde n_{
    b;f'}\rangle\simeq\frac{\tau}{2}\delta_{ff'}\delta_{ab}P_{n_a;f}\;,\\
&\langle \tilde h_{a;f}^*\tilde n_{
    b;f'}\rangle\simeq0\;,
\end{aligned}
    \label{eq:correlator_discrete}
\end{equation}
where $\tau$ denotes the segment length used for the discrete Fourier transform and $\delta_{ff'}$ is a Kronecker delta. 

We then define the cross-correlation estimator $\hat C_f$ for segment $I$ within a frequency bin $[f-\delta f/2,\,f+\delta f/2]$ as~\cite{Allen:1997ad, Romano:2016dpx,Renzini:2023qtj,Zhong:2026owg}
\begin{equation}
    \hat C_f^{(I)}\equiv \frac{20\pi^2f^3}{3H_0^2\tau}\frac{\real\left[\left(\tilde d_{1;f}^{(I)}\right)^*\tilde d_{2;f}^{(I)}\right]}{\gamma_{12;f}}\;.
\end{equation}
Here, $\delta f$ denotes the frequency resolution of the analysis, which can be an integer multiple of $\Delta f\equiv1/\tau$ when the coarse-graining is performed. To simplify the discussion, we consider $\delta f=\Delta f$ in this work.

It is then possible to show that the ensemble average of $\hat C_f$ is equal to $\Omega_{\mathrm{gw};f}$:
\begin{equation}
    \left\langle \hat C_f^{(I)}\right\rangle_{h,\,n}=\frac{10\pi^2f^3}{3H_0^2}P_{\mathrm{gw};f}=\Omega_{\mathrm{gw};f}\;.
    \label{eq:mean_C_f}
\end{equation}
We highlight that the ensemble average in this expression is taken over signal \emph{and} noise realizations, which in practice is equivalent to averaging over a long observation time. To distinguish the various ensemble averages used throughout this work, we append a subscript $x$ to indicate the variable being averaged over: $\langle\cdots\rangle_x \equiv \int (\cdots) p(x) \D x$, where $p(x)$ is the probability density of $x$.

We can also show that the variance of the cross-correlation estimator satisfies the relation~\cite{Zhong:2026owg} 
\begin{equation}
    \left(\hat\sigma_f^{(I)}\right)^2 \propto P_{1;f}P_{2;f}+\gamma_{12;f}P_{\mathrm{gw};f}^2\approx P_{n_1;f}P_{n_2;f}\;,
\end{equation}
where $P_{a;f}\equiv P_{\mathrm{gw};f}+P_{n_a;f}$ and the last approximation is valid in the weak signal regime, where $P_{n_a;f}\gg P_{\mathrm{gw};f}$. The exact expression of $(\hat\sigma^{(I)}_f)^2$ is then given by~\cite{Renzini:2023qtj,Zhong:2026owg}
\begin{equation}
  \left(\hat\sigma^{(I)}_f\right)^2 \simeq \left(\frac{20\pi^2f^3}{3H_0^2}\right)^2 \frac{P_{n_{1};f}^{(I)}P_{n_{2};f}^{(I)}} {8\tau\delta\!f\gamma_{12;f}^2}\;.
  \label{eq:var}
\end{equation}
This result together with Eq.~\eqref{eq:mean_C_f} lays the foundation of the cross-correlation method for detecting a \ac{CGWB}. Assuming a total observation time $T_\mathrm{obs}=N_\mathrm{seg}\tau$, we can average $\hat C_f^{(I)}$ over $N_\seg$ segments to define 
\begin{equation}
    \bar C_f=\frac{1}{N_\seg}\sum_{I=1}^{N_\seg}\hat C_f^{(I)}\;.
\end{equation}
In this case, the expected value of $\bar C_f$ is $\Omega_{\mathrm{gw};f}$, and its variance, $(\bar \sigma_f)^2\equiv (\hat\sigma_f)^2/N_\mathrm{seg}$, is suppressed by the number of segments $N_\mathrm{seg}$. One can define a frequency-dependent \ac{SNR} as $\rho_f\equiv \bar C_f/\bar\sigma_f$ to characterize the detectability of the signal in each frequency bin. Moreover, if the spectral shape of the \ac{SGWB} signal is known \emph{a priori}, one may construct a broadband estimator $\bar C$ with variance $\bar\sigma^2$, and hence a broadband \ac{SNR} $\rho\equiv \bar C/\bar\sigma$. We omit the technical details here and refer interested readers to Refs.~\cite{Romano:2016dpx,Allen:1997ad,Zhong:2026owg} for details of this calculation. Given a sufficiently long $T_\obs$, we can eventually reach an arbitrarily large \ac{SNR} to claim a detection. 

We note that in practice, a weighted average of $\hat C_f^{(I)}$ is performed to account for fluctuations in detector noise over the course of the observation. To simplify the discussion, however, we ignore this complexity and assume the detector noise \ac{PSD} is identical across all segments. 

\subsection{Cross-correlation search for a SGWB of astrophysical origin}
In this subsection, we consider a \ac{SGWB} generated by astrophysical sources. We focus on \ac{CBC} mergers here, although the discussion holds for any kind of sporadic astrophysical sources of \acp{GW}.

Considering $\hat C_f^{(I)}$ defined above, one can further decompose it into four components when the underlying signal is from \acp{CBC}
\begin{widetext}
    \begin{equation}
    \hat C_f^{(I)}=n_f^{(I)}+\bm{1}(N^{(I)}=0)\times 0+\bm{1}(N^{(I)}\geqslant 1)\hat C_f^{(I);\,\mathrm{diag}}+\bm{1}(N^{(I)}\geqslant 2)\hat C_f^{(I);\,\mathrm{off-diag}}\;,
    \label{eq:expansion}
\end{equation}
\end{widetext}
where $n_f^{(I)}$ denotes the detector noise contribution to the cross-correlation estimator, whereas $\tilde n_{a;f}$ in Eq.~\eqref{eq:d_h_n} is the frequency-domain noise strain in detector $a$ in the frequency bin $[f-\delta \!f/2,f+\delta\!f/2]$. $\bm 1(\cdot)$ is the indicator function, which equals one when the condition is satisfied and zero otherwise, and $N^{(I)}$ denotes the number of signals in the segment $I$. The second term corresponds to the case where no \ac{CBC} chirp passes through the time-frequency pixel labeled by $I$ and $f$. The third term captures the contribution from individual events whose chirps intersect this pixel ignoring the interference between different events. The last term encodes the interference (mutual-interaction) of multiple events; this term is non-zero only when multiple chirps overlap in the time-frequency domain and fall within the same time-frequency bin. If two signals both pass through the same time segment but oscillate in different frequency bins, this term is strictly zero. Considering the latest estimate of the \ac{CBC} merger rates~\cite{LIGOScientific:2026ctl}, the probability $p$ of having $N=1$ merger in a $\tau=4$\,s segment is far less than 1 in the sensitive frequency band of current instruments, making the probability of multiple overlapping chirps highly negligible. We therefore ignore the $\Coffdiag$ term in this work, and only consider $\hat C_f^{(I);\,\mathrm{diag}}$. We emphasize a caveat to this approximation: while it is safe for \ac{BBH} and \ac{NSBH} systems, it may break down for \ac{BNS} systems in the low-frequency band below $10$\,Hz, relevant for future instruments. We provide a more detailed discussion of this aspect in Sec.~\ref{subsec:approx_overlap_validity}.
Focusing thus on $\Cdiag$, we can rewrite this term as
\begin{equation}
    \Cdiag=\left\{
    \begin{aligned}
        &\sum_{i=1}^{N^{(I)}}X_f^{(i;I)}+n_f^{(I)},\,&\qquad N^{(I)}\geqslant 1\;,\\
         &n_f^{(I)},\,&\qquad N^{(I)}=0\;.
    \end{aligned}
    \right.
    \label{eq:Cdiag}
\end{equation}
Here, $X_f^{(i;I)}$ denotes the contribution to $\Cdiag$ from the $i$\textsuperscript{th} event crossing this segment within a specific frequency bin. A caveat of our analysis is that, in practice, a single event can extend over more than a single segment in the low-frequency band (self-interaction), therefore the power emitted by that event at a single frequency bin can be separated into multiple segments. We neglect this complexity in this work, and leave this for future investigation. In Sec.~\ref{subsec:approx_overlap_validity} we provide a quantitative estimate of the probability that different kinds of events emit at the same frequency over multiple segments as a function of frequency.
Inserting Eq.~\eqref{eq:Cdiag} into Eq.~\eqref{eq:expansion}, one can then rewrite $\bar C_f$ as
\begin{equation}
    \!\!\!\bar C_f\simeq\frac{1}{N_\seg}\sum_{I=1}^{N_\mathrm{seg}}\left\{
    \begin{aligned}
         &\sum_{i=1}^{N^{(I)}}X_f^{(i;I)}+n_f^{(I)},\,&\quad N^{(I)}\geqslant 1\;,\\
         &n_f^{(I)},\,&\quad N^{(I)}=0\;.
    \end{aligned}
    \right.
\end{equation}
By moving the sum over the events rather than the segments we then obtain
\begin{equation}
\begin{aligned}
    \bar C_f&\simeq\frac{1}{N_\seg}\sum_{i=1}^{N_\tot}X_f^{(i)}+\frac{1}{N_\seg}\sum_{I=1}^{N_\seg}n_f^{(I)}\\
    &\equiv\bar X_f+\bar n_f\;.
\end{aligned}
\label{eq:resum}
\end{equation}
In this form, it is evident that $N_{\tot}$ is a Poisson random variable, whose mean is equal to the product of the total number of segments and the expected number of events per segment, i.e., $N_{\tot}\sim\mathrm{Poisson}(N_{\seg} \mathcal{R}\tau)$, where $\mathcal{R}$ denotes the merger rate in the detector frame. 
Moreover, $X_f^{(i)}$ is independent of $N_\tot$, but depends only on the source parameters of the event, namely $\bm\theta^{(i)}=(m_1,\,m_2,\,z,\,\mathrm{RA}\,,\mathrm{Dec}\,,\psi,...)^{(i)}$, where $m_1$ and $m_2$ denote the primary and secondary masses, $z$ the source redshift, RA and Dec the right ascension and declination, and $\psi$ the \ac{GW} polarization angle. We note that $\bar X_f$ is then a \emph{compound Poisson random variable}. A convenient property of the compound Poisson variable is that its $n$\textsuperscript{th} cumulant $\kappa_n$ is determined solely by its Poisson rate together with the $n$\textsuperscript{th} moment of the individual $X_f^{(i)}$. We therefore have

\begin{align}
\boxed{
\begin{aligned}
\kappa_n(\bar C_f)&=\kappa_n(\bar X_f)+\kappa_n(\bar n_f)\\
&\simeq\frac{\langle N_\tot\rangle}{N_\seg^{n}}\left\langle \left(X_f^{(i)}\right)^n\right\rangle_{\bm\theta}+\kappa_n(\bar n_f)\;.    
\end{aligned}
}
\label{eq:kappa_n}
\end{align}

This equation is the key starting point of our subsequent analysis and serves as a first-order approximation of the cumulants of the cross-correlation estimator when the interference between different chirps and the long time spent in band for \ac{CBC} chirps in the low-frequency band are neglected. In this work, we consider cumulants of $\bar C_f$ up to $n=4$. The first two, $\kappa_1 (\bar C_f)$ and $\kappa_2(\bar C_f)$, are the mean and variance of the $\bar C_f$ distribution, while $\kappa_3(\bar C_f)$ and $\kappa_4(\bar C_f)$ are closely related to its skewness and kurtosis, encoding the non-Gaussianity of the distribution. We explore them in detail in Sec.~\ref{sec:numerical_evaluation}.

The exact expression of $X_f^{(i)}$ is given by
\begin{equation}
    X_f^{(i)}\equiv \frac{20\pi^2f^3}{3H_0^2\tau}\frac{\real[(\tilde h_{1;f}^{(i)})^*\tilde h_{2;f}^{(i)}]}{\gamma_{12;f}}\;,
\end{equation}
and $\tilde h^{(i)}_{a;f}$ with $a\in\{1,2\}$ can be written as
\begin{equation}
    \tilde h^{(i)}_{a;f}=\left(F_{a;f}^+\tilde h_{+;f}^{(i)}+F_{a;f}^\times\tilde h_{\times;f}^{(i)}\right)e^{-\mathrm{i}2\pi f{\hat\Omega\cdot\bm x_a}/{c}}\;,
\end{equation}
where $F_{a;f}^{A}$ with $A\in\{+,\times\}$ denotes the antenna pattern function for detector $a$ and polarization mode $A$, $\hat\Omega$ specifies the sky location of the \ac{GW} source, and $\bm x_a$ is the position vector of detector $a$.

The quantity $\real[(\tilde h_{1;f}^{(i)})^*\,\tilde h_{2;f}^{(i)}]$ can be conveniently expressed in terms of its four Stokes parameters:
\begin{equation}
\begin{aligned}
    I_f &\equiv |\tilde{h}_{+;f}|^2 + |\tilde{h}_{\times;f}|^2 &\; \text{(total intensity)}\;,\\
    Q_f &\equiv |\tilde{h}_{+;f}|^2 - |\tilde{h}_{\times;f}|^2 &\; \text{(linear polarization)}\;,\\
    U_f &\equiv 2\,\real\left[\tilde{h}_{+;f}^*\,\tilde{h}_{\times;f}\right] &\; \text{(linear polarization, rotated)}\;,\\
    V_f &\equiv 2\,\imag\left[\tilde{h}_{+;f}^*\,\tilde{h}_{\times;f}\right] &\; \text{(circular polarization)}\;.
\end{aligned}
\end{equation}
As a side remark, we note that Boybeyi and Mandic adopted a similar Stokes decomposition in a search for an anisotropic \ac{SGWB}~\cite{Boybeyi:2026cvo}, and showed that their framework can probe the contributions from other polarizations that are invisible to conventional intensity-only searches.

We then have
\begin{widetext}
\begin{equation}
\begin{aligned}
\real[(\tilde h_{1;f}^{(i)})^*\,\tilde h_{2;f}^{(i)}]\simeq&\Bigg[\frac{1}{2}(F_1^+F_2^++F_1^\times F_2^\times)^{(i)}\cos\phi_f \cdot I_f^{(i)}-\frac{1}{2}(F_1^+F_2^\times-F_1^\times F_2^+)^{(i)}\sin\phi_f\cdot V_f^{(i)}\\
&+\frac{1}{2}(F_1^+F_2^\times+F_1^\times F_2^+)^{(i)}\cos\phi_f\cdot U_f^{(i)}+\frac{1}{2}(F_1^+F_2^+-F_1^\times F_2^\times)^{(i)}\cos\phi_f\cdot Q_f^{(i)}\Bigg]\\
&\equiv\left[\mathcal{G}_I^{(i)}\cdot I_f^{(i)}+\mathcal{G}_V^{(i)}\cdot V_f^{(i)}+\mathcal{G}_U^{(i)}\cdot U_f^{(i)}+\mathcal{G}_Q^{(i)}\cdot Q_f^{(i)}\right]\\
&=\underbrace{\langle \mathcal{G}_I\rangle _{\hat\Omega}I_f^{(i)}}_{\text{Mean~intensity}}+\underbrace{\left(\mathcal{G}_I^{(i)}-\langle \mathcal{G}_I\rangle_{\hat\Omega}\right)I_f^{(i)}}_{\text{Geometrical shot noise}}+\underbrace{\left(\mathcal{G}_V^{(i)}\cdot V_f^{(i)}+\mathcal{G}_U^{(i)}\cdot U_f^{(i)}+\mathcal{G}_Q^{(i)}\cdot Q_f^{(i)}\right)}_{\text{Polarization leakage}}\;,
\end{aligned}
\label{eq:decomposition}
\end{equation}
\end{widetext}
where we have defined $\phi_f\equiv 2\pi f\hat\Omega\cdot(\bm{x}_1-\bm{x}_2)/c$. 

We identify the first term in the last line of Eq.~\eqref{eq:decomposition} as the ``mean intensity'', as one can show that it is related to the sky-averaged \ac{ORF}
\begin{align}
\langle\mathcal{G}_I\rangle_{\hat\Omega}&=\int\frac{\D\hat\Omega}{4\pi}\frac{1}{2}\left[F_1^+(\hat\Omega)F_2^+(\hat\Omega)+F_1^\times(\hat\Omega)F_2^\times(\hat\Omega)\right]\cos\phi_f\nonumber\\
&=\frac{1}{5}\gamma_{12;f}\;,
\end{align}
and $I^{(i)}$ represents the intensity of the \ac{GW} signal emitted by the $i$\textsuperscript{th} event. As we will show in the following, this term naturally yields the standard expression for the energy density of the \ac{SGWB} generated by \ac{CBC} mergers, $\Omega_f^\mathrm{CBC}$. The second term accounts for the ``geometrical shot noise'' present because the true detector response to individual \ac{CBC} events will deviate from its average value $\langle\mathcal{G}_I\rangle_{\hat\Omega}$. Finally, the third term encodes leakage from other polarization modes, which are present due to finite catalog size. The expectation value of this expression over different realizations of the Universe yields~\cite{Belgacem:2024ohp} 
\begin{equation}
\begin{aligned}
 \langle\real[(\tilde h_{1;f}^{(i)})^*\,\tilde h_{2;f}^{(i)}]\rangle_{\bm\theta}&=\langle \mathcal{G}_I\rangle_{\hat\Omega}\langle I_f\rangle_{\bm\theta}\\
 &=\frac{1}{5}\gamma_{12;f}\left\langle|\tilde{h}_{+;f}|^2 + |\tilde{h}_{\times;f}|^2\right\rangle_{\bm\theta}\;.
\end{aligned}
\end{equation}
From this one obtains a well-known expression for the energy density of the \ac{SGWB} generated by \ac{CBC} events, namely $\Omega_f^\mathrm{CBC}$~\cite{Zhou:2022nmt, Zhou:2022otw,Belgacem:2024ohp,Song:2024pnk, Renzini:2024pxt,Ebersold:2025izh}
\begin{equation}
    \langle\bar C_f\rangle\!=\!\frac{4\pi^2f^3}{3H_0^2}\frac{\langle N_\tot\rangle}{T_\obs}\left\langle|\tilde{h}_{+;f}|^2 \!+\! |\tilde{h}_{\times;f}|^2\right\rangle_{\bm\theta}\equiv\Omega_f^\mathrm{CBC}\;.
    \label{eq:Omega}
\end{equation}
While the interference across different events is neglected in this derivation, it has been shown that this expression holds even when the mutual-interaction is accounted for~\cite{Belgacem:2024ohp}. Therefore, Eq.~\eqref{eq:Omega} remains general, subject only to corrections arising from self-interaction terms.

In the \ac{PTA} literature, Lamb \textit{et al.} initially investigated the non-Gaussianity of $\Omega_f^\mathrm{CBC}$ arising from a finite number of sources~\cite{Lamb:2024gbh}, before shifting their focus to the statistics of the actual \textit{observed} quantities rather than the theoretical $\Omega_f^\mathrm{CBC}$ itself~\cite{Lamb:2025niq}. In the context of LIGO band \ac{SGWB} searches, several studies have explored similar non-Gaussian and shot-noise effects. To the best of our knowledge, Ref.~\cite{Meacher:2014aca} first analyzed the impact of the shot noise on the broadband estimator $\bar C$ as opposed to the narrow-band spectrum. Later, Jenkins \textit{et al.} studied shot noise in directional \ac{SGWB} searches~\cite{Jenkins:2019nks,Jenkins:2019uzp,Kouvatsos:2023bgd}, though their primary focus was constructing an unbiased estimator for the angular power spectrum of the \ac{SGWB} sky. Zhong and Mandic examined the finite-observation-time distribution of $\Omega_f^\mathrm{CBC}$ for \ac{BNS} systems and proposed a method to mitigate its non-Gaussianity~\cite{Zhong:2025gxk}.

However, paralleling the recent shift in PTA literature, we emphasize that the theoretical $\Omega_f^\mathrm{CBC}$ is \textit{not} the quantity our estimator $\bar C_f$ directly measures. Instead of isolating the pure intensity $I_f$, $\bar C_f$ can ``hear'' all polarizations from the \ac{CBC} populations. Consequently, it is subject to geometrical shot noise and polarization leakage, which only average to zero in the limit of sufficiently long observation time or a very large number of sources. While the expected value of $\bar C_f$ is unaffected by the geometrical shot noise and polarization leakage, its higher-order cumulants remain highly sensitive to their contribution. To quantify this, the following section presents numerical results for the cumulants $\kappa_n(\bar C_f)$ up to $n=4$, explicitly detailing the contribution of each term in Eq.~\eqref{eq:decomposition}.

Before discussing our simulations, we clarify an important conceptual distinction. In searches for a \ac{CGWB} using the cross-correlation method, one commonly assumes the \textit{weak-signal regime}, which requires the \ac{SGWB} signal to be a genuinely stochastic process--one that cannot be characterized by a deterministic waveform $h(t;f)$. The scenario considered here is fundamentally different. \ac{CBC} signals interact with our detectors in a deterministic fashion; the only source of randomness is our inability to predict the exact number of sources and their precise source parameters within a given observation period. Consequently, comparing the \ac{PSD} of \ac{CBC} events directly against the detector noise \ac{PSD} does not straightforwardly indicate whether we are in the weak-signal regime, as that regime is defined in the context of truly stochastic signals, such as a \ac{SGWB} of cosmological origin.

%% file: simulations.tex
\section{Simulations}\label{sec:sim}
In this section, we first discuss our choices for the \ac{CBC} populations of \ac{BBH}, \ac{BNS}, and \ac{NSBH} mergers, their characteristics, and the statistical properties of some quantities of relevance. We then show the simulated cumulants $\kappa_n(\bar C_f)$ for $n=1,2,3,4$ for three types of \ac{CBC} mergers.
\subsection{CBC populations}

Our choices for the \ac{CBC} distributions are based on the latest LVK estimates~\cite{LIGOScientific:2026ctl}. We adopt the local merger rates from \textsc{PixelPop}~\cite{Heinzel:2024jlc} for all \ac{CBC} populations: \ac{BBH} at $R_0^\mathrm{BBH}=18.0^{+4.9}_{-4.4}~\mathrm{Gpc}^{-3}~\mathrm{yr}^{-1}$, \ac{BNS} at $R_0^\mathrm{BNS}=23.4^{+54.7}_{-18.2}~\mathrm{Gpc}^{-3}\mathrm{yr}^{-1}$, and \ac{NSBH} at $R_0^\mathrm{NSBH}=15.9^{+16.9}_{-8.9}~\mathrm{Gpc}^{-3}\mathrm{yr}^{-1}$, where the upper and lower limits correspond to the 90\% credible intervals~\cite{LIGOScientific:2026ctl}.

Concerning \ac{BBH} sources, we assume their masses follow the \textsc{Broken Power-Law + 2 Peaks} distribution, with the maximum-likelihood parameters from Ref.~\cite{LIGOScientific:2026ctl}. In this model, the primary source-frame mass is assumed to follow a broken power-law distribution mixed with two left-truncated Gaussian distributions and smoothed at the left and right edges. The mass ratio distribution is instead modeled as a smoothed power law. We assume a redshift distribution with merger rate following the Madau-Dickinson profile~\cite{Madau:2014bja}
\begin{equation}
    p(z)\propto R_0\frac{(1+z)^{\alpha_z-1} [(1+z_p)^{\alpha_z+\beta_z} + 1]}{(1+z)^{\alpha_z+\beta_z} + (1+z_p)^{\alpha_z+\beta_z}} \dfrac{\D V_{\rm c}}{\D z}(z)\;,
    \label{eq:MD}
\end{equation}
where $\alpha_z=2.85,\,\beta_z=4.14,\,z_p=2.27$, corresponding to the maximum likelihood values from Ref.~\cite{LIGOScientific:2026ctl}. The mass and spin distributions are assumed to be independent. We further assume the binaries to have vanishing spins, as their contribution to the overall \ac{SGWB} is negligible~\cite{Ebersold:2025izh}.

For \ac{BNS} sources, we instead assume an uncorrelated uniform distribution for the source-frame component masses $m_1,\,m_2\sim\mathcal{U}[1,\,2]\,{\rm M}_\odot$~\cite{KAGRA:2021duu,Landry:2021hvl,Zhong:2025qno}, vanishing spins. To model the merger rate history, we assume that \ac{BNS} progenitors follow the Madau-Dickinson profile with $\alpha_z=2.6,\,\beta_z=3.6,\,z_p=2.2$~\cite{Madau:2016jbv}. We then convolve it with a power-law time delay distribution between binary formation and merger $P(t_d)\propto t_d^{-1}$ with a minimum time delay of 20\,Myr and a maximum corresponding to the Hubble time; see e.g. Ref.~\cite{Regimbau:2012ir}. We further assume vanishing tidal deformabilities for the objects, as their effect would mostly impact the \ac{SGWB} at high frequencies, where the sensitivity of the detectors is low.

For \ac{NSBH} mergers, we use the \textsc{NSBH-pop} model~\cite{Biscoveanu:2022iue}, which features a truncated
power-law distribution for the source-frame BH mass and a truncated Gaussian distribution for the mass ratio. The parameters are fixed to the maximum-likelihood values from Ref.~\cite{LIGOScientific:2024elc}. We assume vanishing spins for the objects and vanishing tidal deformability for the NSs. The merger rate model is identical to that of the \ac{BNS} population, aside from a different local rate. 

All the remaining extrinsic parameters (sky location, cosine of the inclination, polarization angle, merger time and phase) are sampled uniformly, and the signal originating from each merger is simulated through the \textsc{IMRPhenomXAS} waveform model~\cite{Pratten:2020fqn}, which includes only the dominant quadrupole emission mode. As shown in Ref.~\cite{Renzini:2024pxt}, incorporating higher-order modes in the waveform has a negligible effect on the resulting $\Omega_f^{\rm CBC}$. We therefore leave a more detailed investigation involving non-zero spins, tidal deformabilities, and multiple waveform models for future studies.

We emphasize that, since current observational constraints on \ac{BNS} and \ac{NSBH} populations are significantly weaker than those for \ac{BBH} systems, the population models adopted in this work are purely representative. While these models will naturally require updating as future detections improve our understanding of the \ac{CBC} demographics, we expect the qualitative conclusions of this work to remain robust even under more accurate population assumptions.

\subsection{Region in which the low-overlap probability approximation holds}\label{subsec:approx_overlap_validity}

From the chosen \ac{CBC} populations we can evaluate the expected number of events per time-frequency bin $\mathcal{R}\tau$ as a function of frequency. At high frequencies, this quantity reduces to what is commonly referred to as the duty cycle in the literature~\cite{Smith:2017vfk,Biscoveanu:2020gds,Kou:2025bhk,Lawrence:2023buo}. At low frequencies, however, a single chirp can extend across multiple segments, so $\mathcal{R}\tau$ is no longer determined by the merger rate alone but also depends on how rapidly the chirp frequency evolves with time.

\begin{figure*}
    \includegraphics[width=0.43\linewidth]{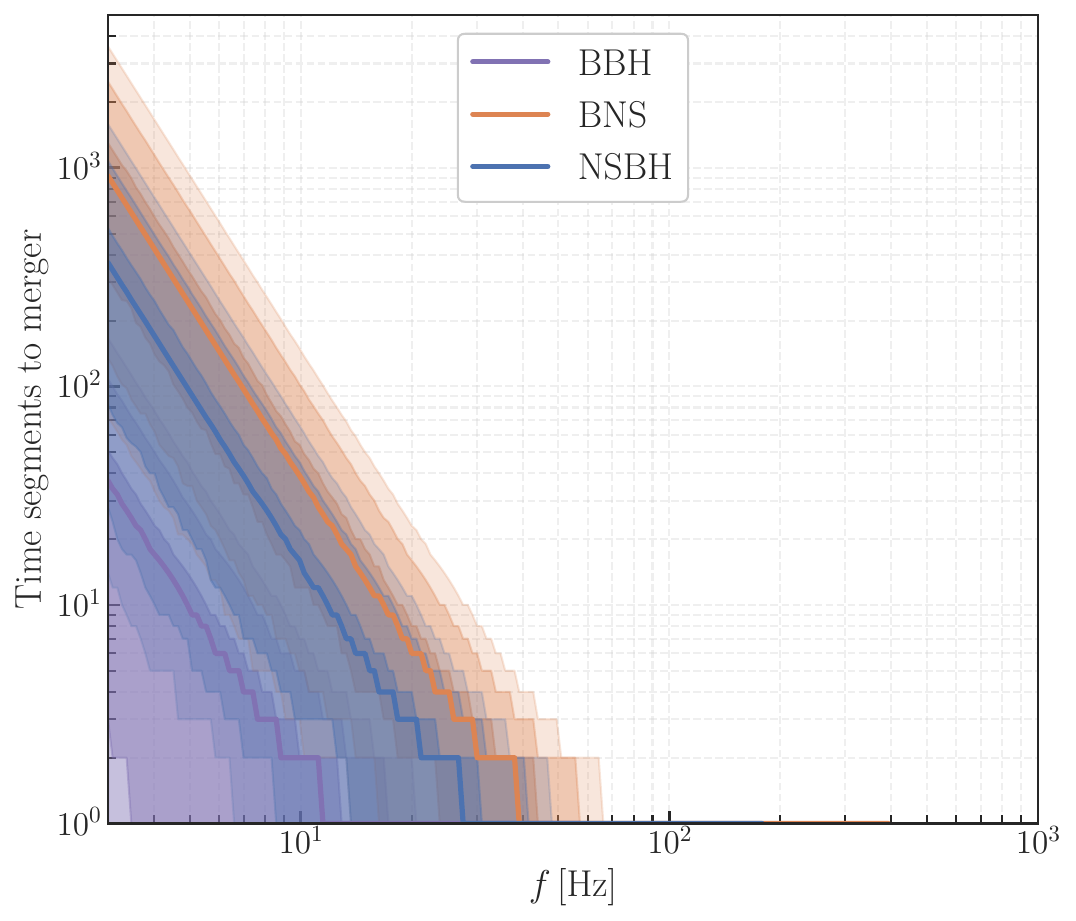} \hfill \includegraphics[width=0.43\linewidth]{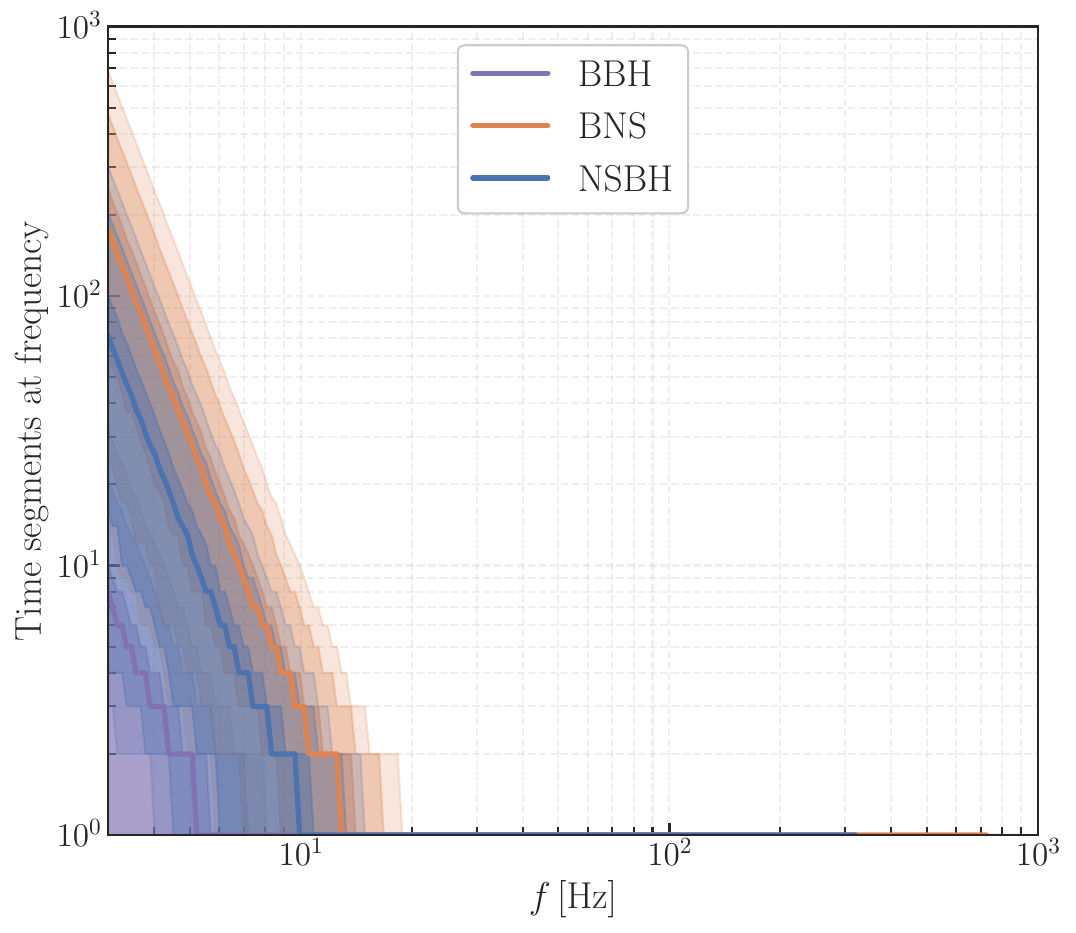}
    \caption{Distributions of the number of time segments occupied until merger (\emph{left panel}) and within a single frequency bin (\emph{right panel}) for the chosen population of \acp{CBC}. The solid lines show the median of the distribution, while the shaded areas the 68\%, 95\%, and 99\% \ac{HDI} with shades of decreasing opacity.}
    \label{fig:nsegm_freq}
\end{figure*}

We start by evaluating the distribution of the number of time segments occupied by an event from a given frequency to merger and within a single frequency bin. The latter is obtained by computing the difference in time to merger at the edges of adjacent frequency bins, and converting this to a number of time segments with the chosen $\tau$. 
We recall that in our analysis $\tau=4~\mathrm{s}=1/\delta f$. The results are shown in Fig.~\ref{fig:nsegm_freq}: here, for each of the three \ac{CBC} populations, we report the median and 68\%, 95\%, and 99\% \acp{HDI} for these distributions, obtained by sampling $5\times10^5$ events from each of them. The values are obtained using the expression for the time to coalescence as a function of frequency from Ref.~\cite{Buonanno:2009zt}, accurate up to 3.5\,\ac{PN} order for non-spinning binaries.
For \acp{BBH}, the vast majority of the events only occupy one time segment until merger already at $\sim\!10\,$Hz, while \ac{BNS} and \ac{NSBH} systems can occupy several time segments until $\sim\!40\,$Hz. The lower the frequency, the higher the number of segments occupied, which can increase to more than $10^3$ for \acp{BNS} at 3\,Hz. Below $\sim\!10$\,Hz the stationary-phase duration of an event at frequency $f$ exceeds a single segment for \acp{BNS} and \acp{NSBH}, so single events contribute coherently to multiple adjacent segments. 

\begin{figure}[tbp]
    \centering
    \includegraphics[width=0.95\linewidth]{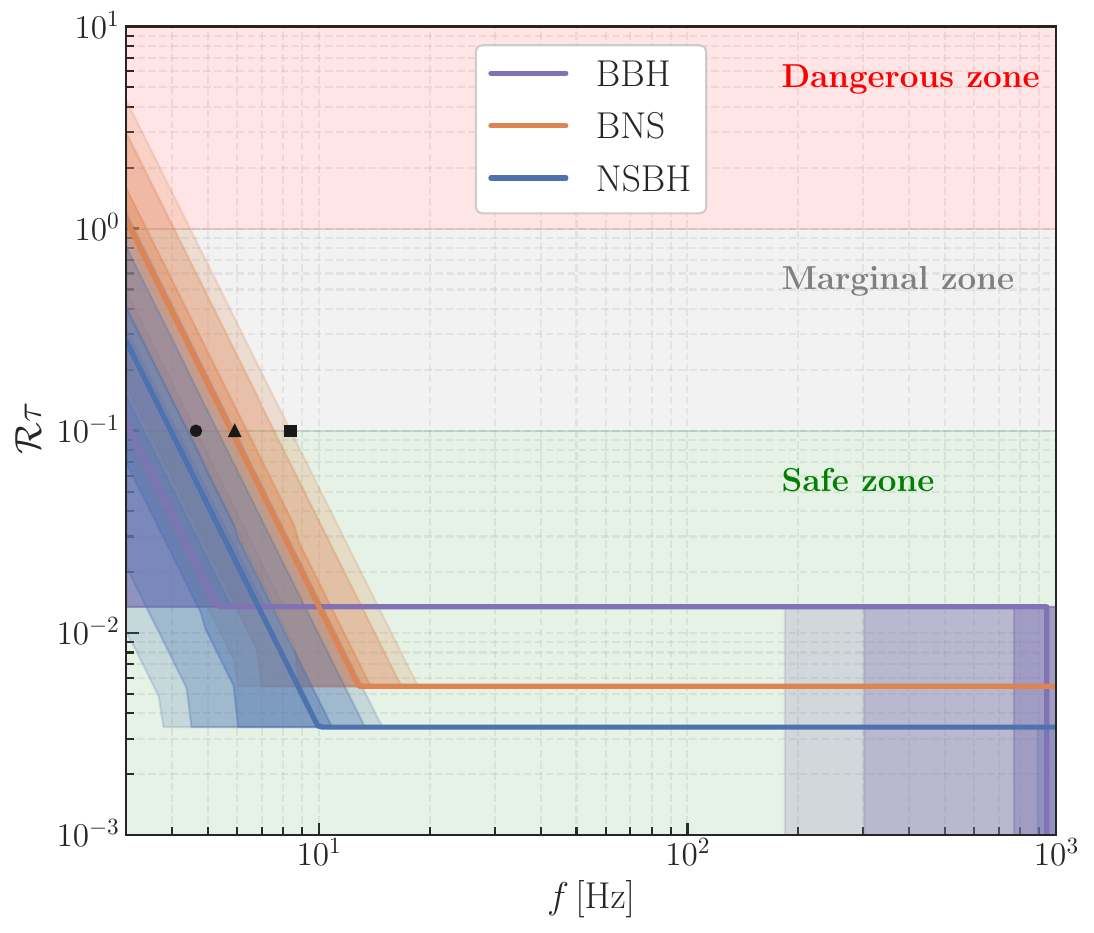}
    \caption{Expected number of events per time-frequency bin $\mathcal{R}\tau$ as a function of frequency for the various \ac{CBC} populations considered. The solid lines show the median of the distribution, while the shaded areas the 68\%, 95\%, and 99\% \ac{HDI} with shades of decreasing opacity. The markers highlight the points in which the upper 99\% \ac{HDI} of each population crosses to the safe zone (circle for \acp{BBH}, triangle for \acp{NSBH}, and square for \acp{BNS}).
    The drop observed in the \ac{BBH} curve at $f\gtrsim150\,{\rm Hz}$ is because a fraction of the events merge at these frequencies, and thus do not occupy any bin.}
    \label{fig:Ratu}
\end{figure}

In Fig.~\ref{fig:Ratu}, we report the distributions of $\mathcal{R}\tau$ obtained for the three \ac{CBC} populations considered.
This is computed as follows: for each event  $i$, the time that this source spends inside a frequency bin centered at $f$ is
\begin{equation}
    \Delta t_{i; f} = \int_{f-\delta f/2}^{f +\delta f/2} \dfrac{{\rm d}f'}{\dot{f}_{\rm gw}^{(i)}(f')} \approx \dfrac{\delta f}{\dot{f}_{\rm gw}^{(i)}(f)} = \dfrac{1}{\tau\dot{f}_{\rm gw}^{(i)}(f)}\;,
\end{equation}
where $\dot{f}_{\rm gw}$ denotes the frequency evolution of the binary as a function of frequency, and the approximation holds for small $\delta f$. The expected number of events occupying that frequency bin at a random time for the chosen time of observation is then 
\begin{equation}
    {\cal R}\tau (f) = \dfrac{1}{T_{\rm obs}} \sum_i^{N_{\rm tot}} \dfrac{1}{\tau\dot{f}^{(i)}_{\rm gw}(f)}\;.
\end{equation}
To compute $\dot{f}_{\rm gw}$ we use a 3.5\,\ac{PN}-accurate expression~\cite{Buonanno:2009zt}, and we employ a large set of $5\times10^5$ samples from each distribution, and rescale to the chosen yearly event rate. The drop observed in the \ac{BBH} distribution at $f\gtrsim150\,{\rm Hz}$ happens because a fraction of the events merge at these frequencies, hence they do not occupy any time-frequency bin.

When $\mathcal{R}\tau\ll1$, the probability of two events overlapping within a single time-frequency bin is negligible, and our assumption of neglecting $\Coffdiag$ in Eq.~\eqref{eq:expansion} is well justified. When $\mathcal{R}\tau\gtrsim\mathcal{O}(1)$, this assumption breaks down and the contribution from $\Coffdiag$ must be carefully accounted for. We therefore adopt the following empirical classification: $\mathcal{R}\tau\geqslant1$ defines the ``dangerous zone'', $0.1\leqslant \mathcal{R}\tau< 1$ the ``marginal zone'', and $\mathcal{R}\tau<0.1$ the ``safe zone''. 


As one can see, \ac{BBH} systems occupy the marginal zone even at very low frequencies, with the 99\% upper \ac{HDI} crossing to the safe zone at $f\sim4.6\,$Hz. For \acp{BNS} the probability of overlap in a time-frequency bin is non-negligible up to higher frequency, with the 99\% upper \ac{HDI} crossing to the safe zone at $f\sim8.4\,$Hz. The \ac{NSBH} distribution sits in between, with the 99\% upper \ac{HDI} crossing to the safe zone at $f\sim5.9\,$Hz. If we use the upper 90\% value of the local merger rate for the three classes of sources reported in Ref.~\cite{LIGOScientific:2026ctl} for the \textsc{PixelPop} model~\cite{Heinzel:2024jlc}, we find the crossings of the 99\% upper \ac{HDI} to happen at $f\sim5.0\,$Hz, $f\sim11.7\,$Hz, and $f\sim7.2\,$Hz for \acp{BBH}, \acp{BNS}, and \acp{NSBH}, respectively. Instead, for the lower 90\% local merger rate the crossings instead happen at frequencies as low as $f\sim4.3\,$Hz, $f\sim5.6\,$Hz, and $f\sim4.7\,$Hz for \acp{BBH}, \acp{BNS}, and \acp{NSBH}, respectively. We thus expect our two approximations of neglecting self and mutual-interactions to hold at $f\gtrsim5\,$Hz for \acp{BBH}, $f\gtrsim12\,$Hz for \acp{BNS}, and $f\gtrsim7\,$Hz for \acp{NSBH}.  


\subsection{Numerical evaluation of \texorpdfstring{$\kappa_n(\bar C_f)$}{kappa\_n(C\_f)}}\label{sec:numerical_evaluation}

To compute $\kappa_n(\bar C_f)$, we must numerically evaluate ensemble averages of the form $\langle (X_f^{(i)})^n\rangle_{\bm\theta}$. For $n=1$, this yields the usual energy density spectrum: $\kappa_1(\bar C_f)_{\bm\theta}=\langle\bar C_f\rangle=\Omega_f^\mathrm{CBC}$. In the literature, this quantity is typically evaluated via \ac{MC} integration~\cite{Renzini:2024pxt,Ebersold:2025izh}. In this approach, one draws $N_\mathrm{draw}\gg 1$ events from the underlying population model and approximates the expectation value $\langle X_f^{(i)}\rangle_{\bm\theta}$ as
\begin{equation}
    \left\langle X_f^{(i)}\right\rangle_{\bm\theta}\simeq\frac{1}{N_\mathrm{draw}}\sum_{i=1}^{N_\mathrm{draw}}X_f^{(i)}(\bm\theta)\;.
\end{equation}

This approach may fail when the distribution of the quantity being averaged has sharp and narrow features or is heavy-tailed. In our analysis, $X_f^{(i)}$ captures the detector response to the \ac{GW} power, and the power emitted, e.g., by a nearby and massive system can be higher than that of a distant and low-mass system by several orders of magnitude. In this case, the mean of $(X_f^{(i)})^n$ is primarily dominated by large values of $X_f^{(i)}$. If the generated $X_f^{(i)}$ sample is incomplete or does not sufficiently probe the large $X_f^{(i)}$ region, the resulting $\langle(X_f^{(i)})^n\rangle$ can be significantly biased relative to the true value. Moreover, evaluating the numerical uncertainty of \ac{MC} estimators requires additional \ac{MC} calculations, which face similar issues. Therefore, both the estimator itself and its \ac{MC} uncertainties may be \textit{unreliable}. To avoid this issue and increase the stability of our estimates, we use the \href{https://github.com/gplepage/vegas}{\textsc{vegas}} Python package~\cite{Vegas,Lepage:1977sw,Lepage:2020tgj} to perform the numerical \ac{MC} integration. 

\begin{figure*}[!htbp]
    \centering
\includegraphics[width=0.85\linewidth]{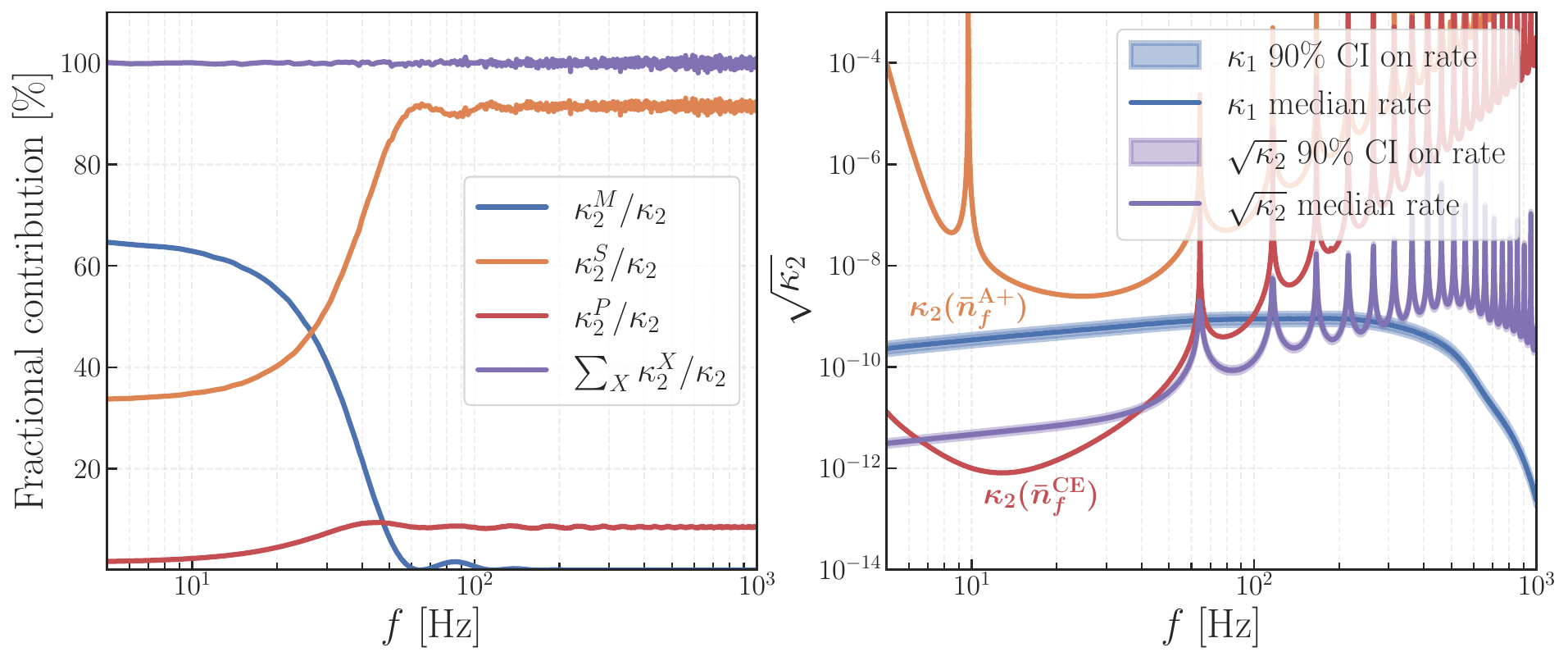}
    \caption{Statistical properties of the cross-correlation estimator $\bar C_f$ for the \ac{SGWB} from a \ac{BBH} population. Left: fractional contributions of the mean intensity ($\kappa_2^M$, blue), geometrical shot noise ($\kappa_2^S$, orange), and polarization leakage ($\kappa_2^P$, red) to the total $\kappa_2(\bar X_f)$. The purple curve shows their sum, $\sum_X\kappa_2^X/\kappa_2$, which remains close to 100\% across the full frequency band, confirming that cross-term contributions are negligible. Right: comparison of $\kappa_1(\bar C_f)=\Omega_f^\mathrm{BBH}$ (blue) with $\sqrt{\kappa_2(\bar X_f)}$ (purple) and $\sqrt{\kappa_2(\bar n_f)}$ for two detector configurations: two LIGO detectors at A+ sensitivity (orange) and two \ac{CE} detectors with their best forecasted sensitivity (red). All curves assume one year of observation with segment length $\tau=4$ s, and frequency resolution $\delta f=1/4$\,Hz. The shaded regions represent the range of $\kappa_1(\bar C_f)$ and $\sqrt{\kappa_2(\bar X_f)}$ corresponding to the 90\% credible interval on the local merger rate.}
    \label{fig:BBH_kappa2}
\end{figure*}

Finally, we note that the theoretical expectation value $ \langle (X_f^{(i)})^n\rangle_{\bm\theta}$ may diverge if the minimum redshift is set to $z_\mathrm{min}=0$, as $X_f$ diverges for a source at $z=0$~\cite{Jenkins:2019uzp,Zhong:2025gxk}. To guarantee the convergence of these moments, we impose a cutoff at $z_{\min}=0.2$. The choice of $z_{\min}$ is somewhat arbitrary and does not affect the main conclusions of this work; we refer the reader to Ref.~\cite{Zhong:2025gxk} for a detailed discussion. It could in principle be optimized using a suitable figure of merit, but we defer this to future work. The qualitative conclusions of this work remain robust against the exact choice of this lower bound, provided $z_\mathrm{min}$ remains small, below $\mathcal{O}(0.5)$. Choosing a large $z_\mathrm{min}$ means removing a great number of events, potentially yielding a non-negligible residual background and increasing the non-Gaussianity of the remaining signal. 
The removal of resolvable events is imperfect; as a result, the signals that survive this procedure form a residual background.
In Ref.~\cite{Zhong:2025gxk}, it was shown that the event-removal residual remains minimal for $z_\mathrm{min}=0.35$, whereas the residual associated with larger values of $z_\mathrm{min}$ can become non-negligible. 
\subsection{Detector choices}\label{sec:detector_choice}
We consider two detector networks in this paper. For both networks, we assume the same location and orientation as the two LIGO interferometers, i.e., LIGO Hanford and LIGO Livingston~\cite{LIGOScientific:2014pky}. For the first network, we assume both detectors operate at LIGO A+ design sensitivity~\cite{O5_PSD_curve}, which is the target sensitivity for the fifth LVK observing run (O5). For the second network, we assume both detectors operate at the best current forecasted sensitivity of \ac{CE}~\cite{noise_curves_gwforge}, available in the \href{https://github.com/koustavchandra/gwforge}{\textsc{gwforge}} package~\cite{Chandra:2024dhf}. For both networks, we assume an observation time of one year without any duty cycle, corresponding to $N_\mathrm{seg}=1\,\mathrm{year\,/\,4\,s\sim7.9\times10^{6}}.$
\subsection{BBH background}
In Fig.~\ref{fig:BBH_kappa2} we show results for the \ac{SGWB} generated by the \ac{BBH} population. The left panel shows the fractional contribution of $\kappa_2^{X}(\bar X_f)$ to the full $\kappa_2(\bar X_f)$, where $X\in\{M,\,S,\,P\}$ denotes the mean intensity, geometrical shot noise, and polarization leakage, respectively. We emphasize that both $\kappa_2(\bar X_f)$ and $\kappa_2(\bar n_f)$ contribute to the variance of $\bar C_f$, i.e., $\kappa_2(\bar C_f)$. The variance due to the \ac{CBC} signal itself has usually been neglected in the literature, where only the detector noise is accounted for. Here, however, we carefully study $\kappa_2(\bar X_f)$ to quantify its contribution to the total variance, along with the contribution of each individual term defined in Eq.~\eqref{eq:decomposition}. If the predicted $\kappa_2(\bar X_f)\gtrsim\kappa_2(\bar n_f)$, then the contribution from the signal itself cannot be neglected. As a side remark, we note that $\kappa_2(\bar n_f)$ [cf. Eq.~\eqref{eq:var}] is the variance of the cross-correlation estimator due to detector noise, which is \emph{different} from the power-law integrated sensitivity curve~\cite{Thrane:2013oya} that quantifies the sensitivity of a detector network to detect a power-law \ac{SGWB} signal.

Below 25\,Hz, we find that the mean intensity term contributes more than 60\% of $\kappa_2(\bar X_f)$, but its contribution becomes negligible for $f\gtrsim100$\,Hz. The geometrical shot noise term overtakes the mean intensity above 25\,Hz, and dominates $\kappa_2(\bar X_f)$ throughout the remaining frequency band. The polarization leakage term is negligible below 40\,Hz, and saturates at the $\sim8\%$ level above 40\,Hz. In addition to these three contributions, we also plot their sum (purple curve), which remains close to 100\% across the entire frequency band, indicating that contributions to $\kappa_2$ from the cross-terms are negligible. As a side remark, we note that the cross-terms here do \emph{not} refer to the off-diagonal terms in Eq.~\eqref{eq:expansion}, but instead denote the joint cumulants among $M$, $S$, and $P$.

\begin{figure*}[tbp]
    \centering
\includegraphics[width=0.85\linewidth]{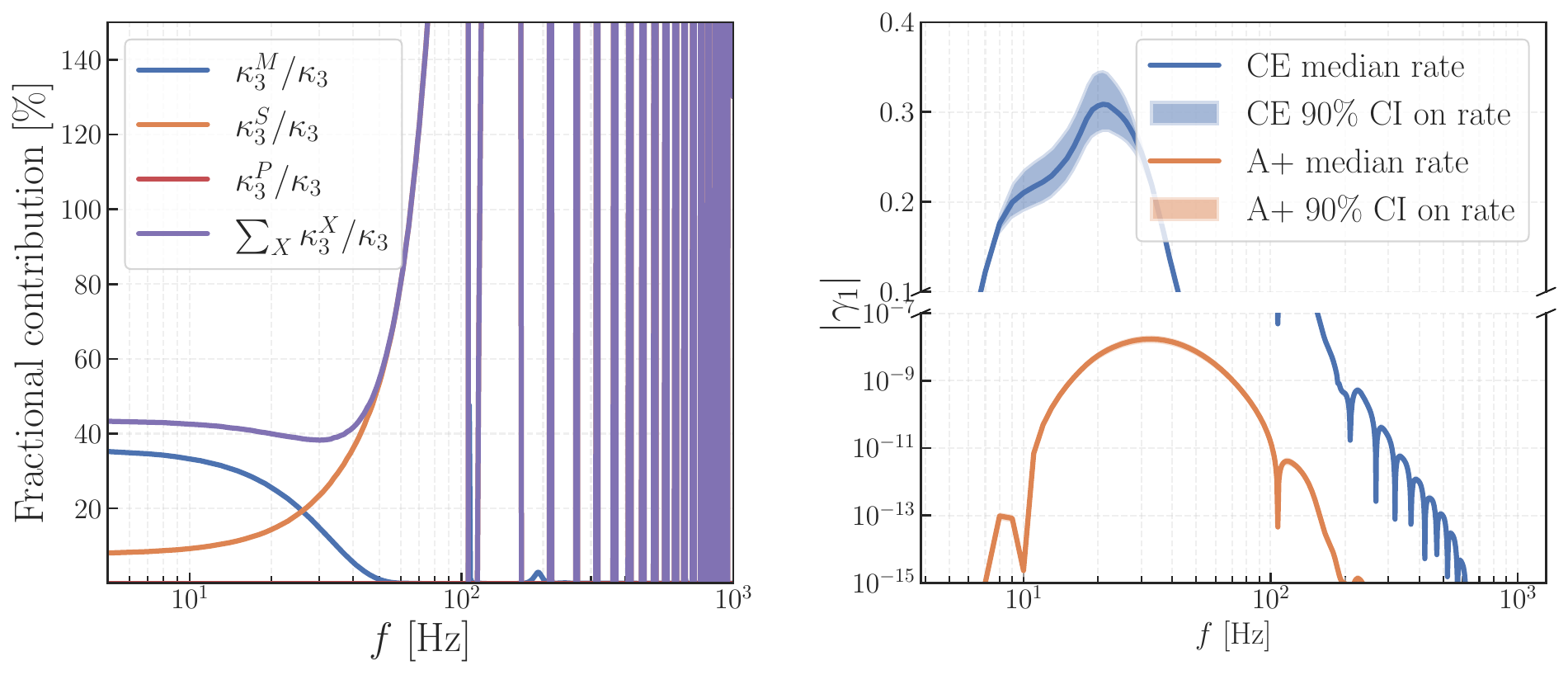}
    \caption{As in Fig.~\ref{fig:BBH_kappa2}, but shows the third-order statistics of $\bar C_f$ for the \ac{BBH} population. Left: the purple curve deviates substantially from 100\%, indicating that cross-term contributions are no longer negligible at third order. The rapid oscillations above $\sim$ 100\,Hz reflect alternating positive and negative contributions. Right: absolute value of the skewness $|\gamma_1(\bar C_f)|$ for the LIGO A+ network (orange) and the CE network (blue), shown with a broken vertical axis.}
    \label{fig:BBH_kappa3}
\end{figure*}
\begin{figure*}[tbp]
    \centering
\includegraphics[width=0.85\linewidth]{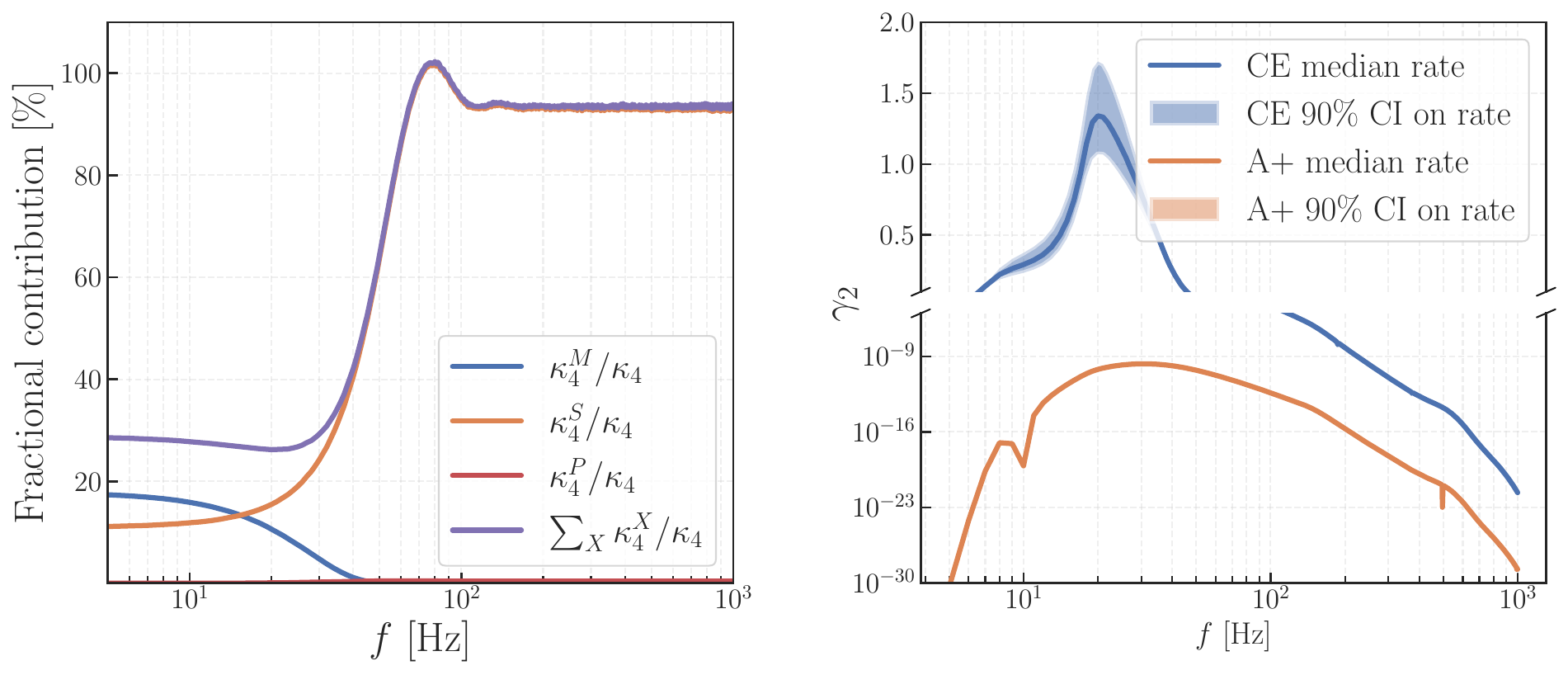}
    \caption{Same as Fig.~\ref{fig:BBH_kappa3}, for the fourth-order statistics of $\bar C_f$. The right panel shows the excess kurtosis $\gamma_2$.}
    \label{fig:BBH_kappa4}
\end{figure*}

In the right panel of Fig.~\ref{fig:BBH_kappa2}, we compare $\kappa_1(\bar C_f)$ (blue), $\sqrt{\kappa_2(\bar X_f)}$ (purple), and $\sqrt{\kappa_2(\bar n_f)}$ (orange and red). The blue and purple solid curves show predictions using the median \ac{BBH} local rate, while the blue and purple bands depict the range of predictions corresponding to the 90\% credible interval on the local rate. Since $\kappa_2(\bar C_f)=\kappa_2(\bar X_f)+\kappa_2(\bar n_f)$, we additionally plot $\sqrt{\kappa_2(\bar n_f)}$ in orange and red. These two curves are computed via Eq.~\eqref{eq:var}, corresponding to the network choice introduced in Sec.~\ref{sec:detector_choice}. The orange curve corresponds to both detectors operating at LIGO A+ sensitivity; the red curve corresponds to both detectors operating at the best projected sensitivity of \ac{CE}.

Comparing the blue and orange curves shows that, in the near future, the detection of the \ac{SGWB} from the \ac{BBH} population will remain the main target of the stochastic background searches. Although the blue curve lies entirely below the noise, the \ac{SGWB} may still be detected by integrating $\bar C_f$ across the full frequency band. Such a detection is expected in the near future~\cite{LIGOScientific:2025bgj}.

Comparing instead the blue and red curves, we see that $\Omega_f^\mathrm{BBH}$ exceeds the detector noise in the most sensitive band of the detector network, making it detectable. 
The purple curve, however, cuts through the red curve in this band, indicating that the variance of $\bar C_f$ is no longer noise-dominated but is instead dominated by the signal itself. While neglecting $\kappa_2(\bar X_f)$ does not affect detection, it would lead to an underestimation of the uncertainty in any subsequent Bayesian analysis of $\bar C_f$ that accounts only for detector noise. We further emphasize that the narrow widths of the blue and purple bands indicate that the uncertainty on the local rate of \ac{BBH} mergers only has a \emph{minimal} impact on $\kappa_1(\bar C_f)$ and $\sqrt{\kappa_2(\bar X_f)}$.

We then show $\kappa_3(\bar X_f)$ in Fig.~\ref{fig:BBH_kappa3}. As in the left panel of Fig.~\ref{fig:BBH_kappa2}, we plot the fractional contributions of each of the three terms to the full $\kappa_3(\bar X_f)$. A similar pattern is observed: $\kappa_3^M$ dominates over the other two terms at low frequencies, while $\kappa_3^S$ takes over above 25\,Hz, and $\kappa_3^P$ is negligible across the entire frequency band. A notable difference, however, is that the cross-terms now contribute significantly to $\kappa_3(\bar X_f)$. Below 25\,Hz, we find $\sum_X\kappa_3^X/\kappa_3^F\lesssim 45\%$, implying that the cross-terms account for the remaining $55\%$. The oscillations seen in the orange and purple curve at high frequencies arise because their contribution alternate between positive and negative values.


The third cumulant $\kappa_3(\bar X_f)$ encodes the \textit{non-Gaussianity} of the estimator $\bar C_f$. Specifically, it measures the asymmetry of the distribution about its mean. A positive $\kappa_3$ indicates a longer right tail; negative $\kappa_3$ a longer left tail. A more useful quantity for characterizing this non-Gaussianity is the skewness $\gamma_{1}$, defined by 
\begin{equation}
    \gamma_{1}(\bar C_f)\equiv\frac{\kappa_3(\bar C_f)}{\kappa_2(\bar C_f)^{3/2}}=\frac{\kappa_3(\bar X_f)}{\left(\kappa_2(\bar X_f)+\kappa_2(\bar n_f)\right)^{3/2}}\;.
\end{equation}

Here we have assumed that the detector noise is Gaussian, so that $\kappa_3(\bar n_f)\equiv0$. We show $|\gamma_1|$ in the right panel of Fig.~\ref{fig:BBH_kappa3}. For the LIGO A+ network, $|\gamma_1|\lesssim\mathcal{O}(10^{-7})$ across the entire frequency band, since $\kappa_2(\bar C_f)\simeq\kappa_2(\bar n_f)\gg\kappa_2(\bar X_f)$ in this regime; the non-Gaussianity of $\bar C_f$ can therefore be safely ignored in the near future. For the \ac{CE} network, however, $\gamma_1=0.31^{+0.04}_{-0.03}$ in the most sensitive band, indicating a non-negligible but moderate skewness. We further note that the peak near $20\,\mathrm{Hz}$ carries \emph{positive} $\gamma_1$ values, consistent with our expectation that the $\bar C_f$ distribution should have a long right tail extending to large values. The alternating sign of $\kappa_3(\bar X_f)$ causes $\gamma_1(f)$ to oscillate around zero at high frequencies, as reflected in the dips visible in the right panel of Fig.~\ref{fig:BBH_kappa3}. Since $|\gamma_1(f)|\ll1$ throughout this regime, however, the resulting skewness is negligible and can safely be ignored.

We show analogous results for $\kappa_4(\bar C_f)$ in Fig.~\ref{fig:BBH_kappa4}. As for to $\gamma_1$, we define the \textit{excess kurtosis} $\gamma_2$ as
\begin{equation}
    \gamma_2(\bar C_f)\equiv \frac{\kappa_4(\bar C_f)}{\kappa_2(\bar C_f)^2}=\frac{\kappa_4(\bar X_f)}{\left(\kappa_2(\bar X_f)+\kappa_2(\bar n_f)\right)^{2}},
\end{equation}
which measures the heaviness of the tail of the distribution relative to a Gaussian distribution. Again, in the above we have used $\kappa_4(\bar n_f)=0$, which follows from the Gaussian noise assumption. We find excess kurtosis $\gamma_2=1.3^{+0.4}_{-0.3}$ for $\bar C_f$ under the \ac{CE} sensitivity at $f=20$ Hz, while $\gamma_2$ remains negligible for the LIGO A+ network. Similarly to Fig.~\ref{fig:BBH_kappa3}, the cross-terms dominate $\kappa_4$ below $f\lesssim50\,\mathrm{Hz}$, and the geometrical shot noise surpasses all other terms and contributes more than 90\% at $f\gtrsim 100\,\mathrm{Hz}.$ Again, the polarization leakage is negligible.
\begin{figure}[tbp]
    \centering
\includegraphics[width=0.85\linewidth]{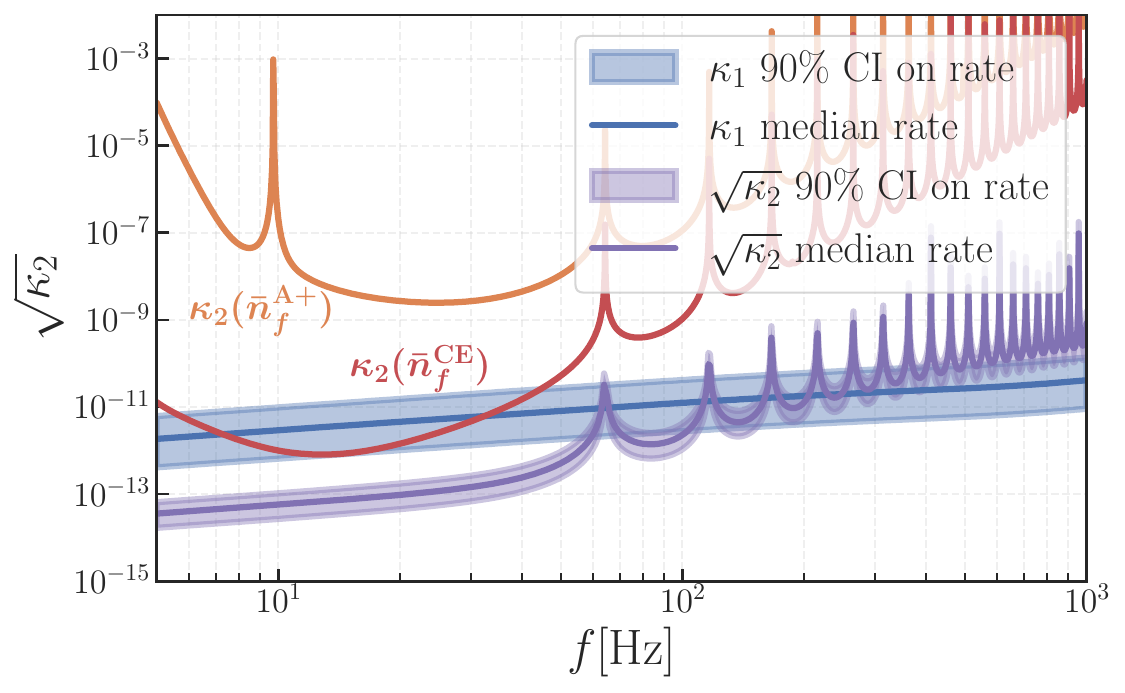}
    \caption{Same as the right panel of Fig.~\ref{fig:BBH_kappa2}, for the \ac{BNS} case.} 
    \label{fig:BNS_kappa12}
\end{figure}

\begin{figure*}[tbp]
    \centering
\includegraphics[width=0.85\linewidth]{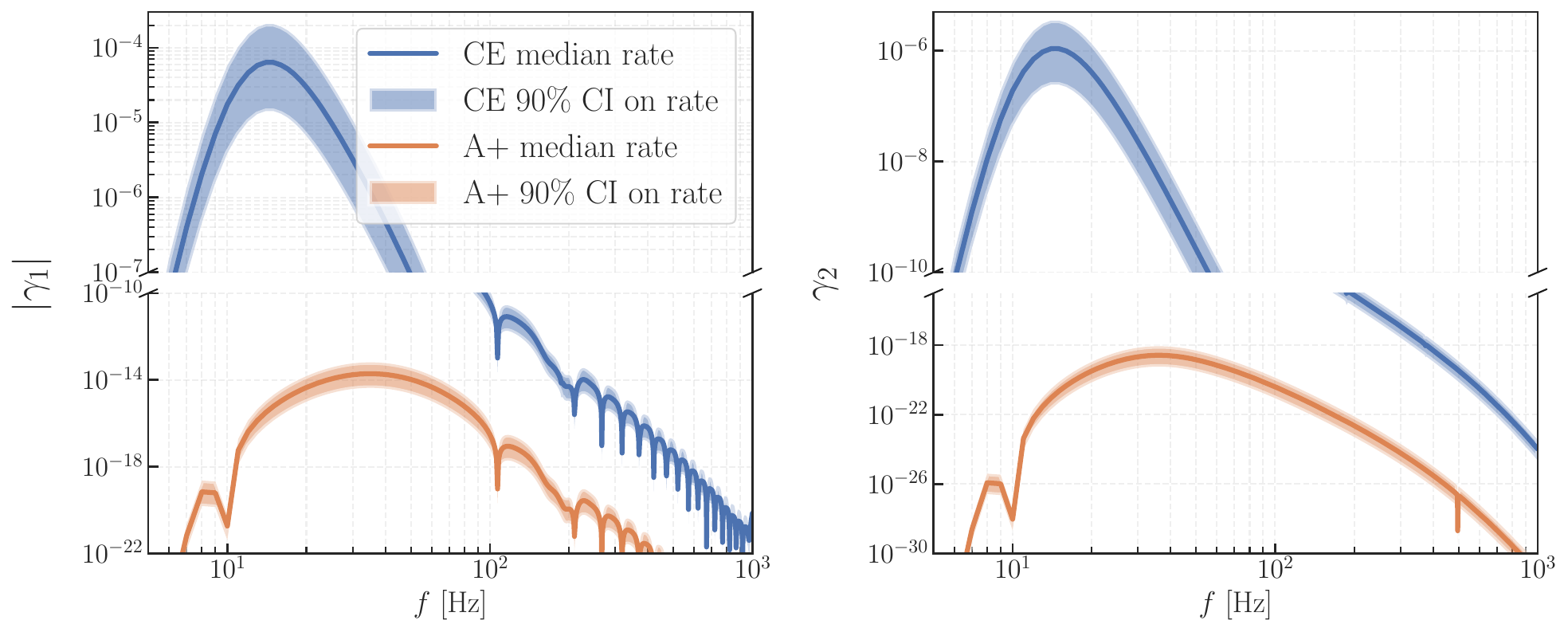}
    \caption{Skewness $|\gamma_1(\bar C_f)|$ (left) and excess kurtosis $\gamma_2(\bar C_f)$ (right) for the \ac{SGWB} from the \ac{BNS} population, shown for the LIGO A+ network (orange) and the \ac{CE} network (blue). Both $|\gamma_1|$ and $\gamma_2$ are negligible in this scenario.}
    \label{fig:BNS_kappa34}
\end{figure*}
We stress that this finding highlights the need for more sophisticated methods to model the non-Gaussianity. Previous works have incorporated weak non-Gaussianity via the Edgeworth expansion~\cite{Martellini:2015mfr,Martellini:2014xia, Liu:2026xhc}. As an asymptotic rather than convergent series, its truncation is formally justified only for small departures from Gaussianity~\cite{Blinnikov:1997jq}, and it can develop regions of negative density in the tails for sufficiently large cumulants. For the values of $|\gamma_1|$ and $\gamma_2$ inferred here (Fig.~\ref{fig:BBH_kappa3} and Fig.~\ref{fig:BBH_kappa4}), we find that the expansion remains positive-definite throughout the bulk of the distribution (a valid probability density must be non-negative everywhere, so positivity is a necessary but not sufficient condition for the reliability of the truncated expansion). The expansion is therefore adequate as a leading-order characterization of the non-Gaussianity, though caution is warranted if the tail behavior is of interest. Beyond developing more refined statistical models, we note that in the \ac{XG} era the \ac{SGWB} from \ac{BBH} systems may no longer be a primary scientific target; in that case, one can simply notch out the resolved \ac{BBH} signals~\cite{Zhong:2022ylh,Zhong:2024dss,Zhong:2025qno} from the data stream and focus on the \ac{SGWB} contributions from other types of \acp{CBC}.
\subsection{BNS and NSBH background}
We now turn to the \ac{SGWB} generated by \acp{BNS}. Since $\kappa_2(\bar X_f)$ is itself smaller than $\kappa_2(\bar n_f)$, we no longer break down its fractional contributions by source. Fig.~\ref{fig:BNS_kappa12} shows the \ac{BNS} analog of the right panel of Fig.~\ref{fig:BBH_kappa2}. In this case, we find $\sqrt{\kappa_2(\bar X_f)}/\sqrt{\kappa_2(\bar n_f)}\leqslant 8.6^{+7.1}_{-4.5}\%$, with the maximum reached at the most sensitive frequency $\sim\! 14\,\mathrm{Hz}.$ In principle, this non-negligible contribution should be properly accounted for in any downstream analyses that rely on $\bar C_f$ and its variance $\kappa_2(\bar C_f)$, to avoid any potential bias. We further show the skewness and excess kurtosis of $\bar C_f$ in the left and right panels of Fig.~\ref{fig:BNS_kappa34}, respectively. The non-Gaussianity is substantially weaker for the \ac{BNS} background than for the \ac{BBH} case for both detector networks considered here, with $|\gamma_1(\bar C_f)|\leqslant 6.2^{+14}_{-5}\times 10^{-5}$ and $\gamma_2(\bar C_f)\leqslant 1.1^{+2.4}_{-0.8}\times 10^{-6}$. Although Fig.~\ref{fig:Ratu} indicates that our approximation [cf. Eq.~\eqref{eq:kappa_n}] may break down for \ac{BNS} below 8.4\,Hz (12\,Hz for the upper 90\% local merger rate from Ref.~\cite{LIGOScientific:2026ctl}), we expect the conclusion of negligible non-Gaussianity to remain robust, unless the error induced by the two approximations we adopt reaches $\mathcal{O}(10^4)$, which we consider unlikely.

\begin{figure}[tbp]
    \centering
\includegraphics[width=0.85\linewidth]{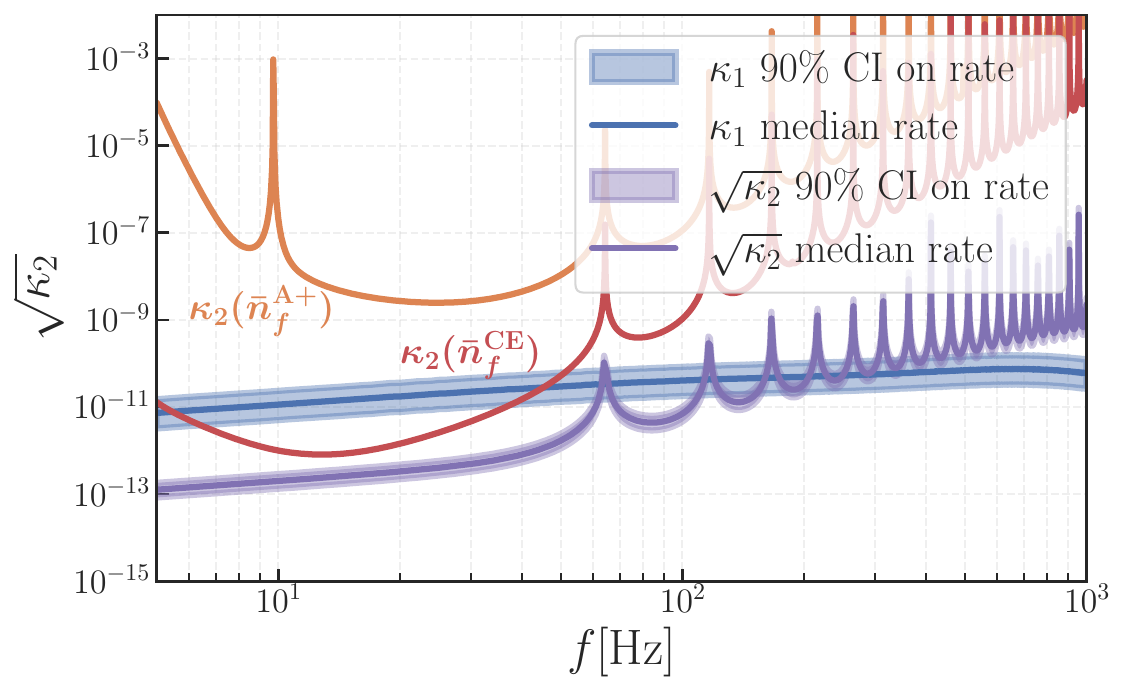}
    \caption{Same as Fig.~\ref{fig:BNS_kappa12}, for the \ac{NSBH} case.}
    \label{fig:NSBH_kappa12}
\end{figure}

\begin{figure*}[tbp]
    \centering
\includegraphics[width=0.85\linewidth]{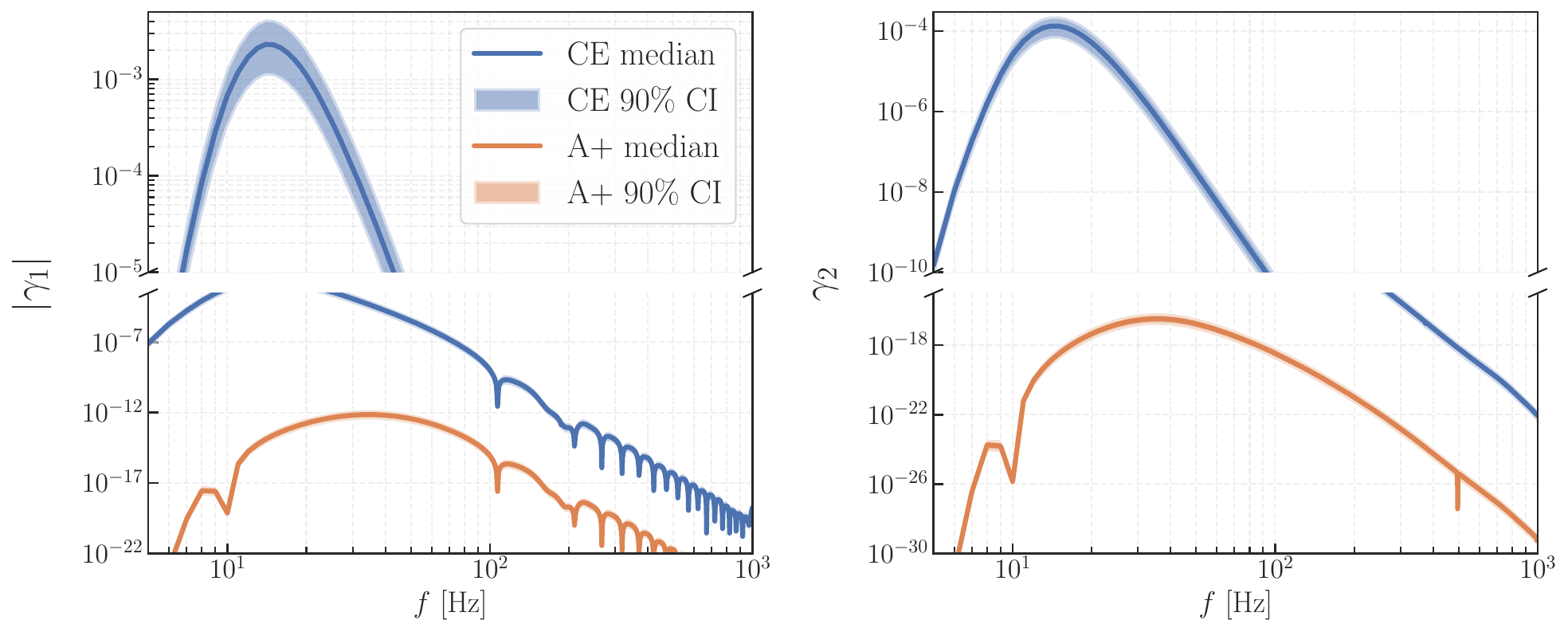}
    \caption{Same as Fig.~\ref{fig:NSBH_kappa12}, for the \ac{NSBH} case.}
    \label{fig:NSBH_kappa34}
\end{figure*}

We similarly present the numerical results for the \ac{NSBH} case in Fig.~\ref{fig:NSBH_kappa12} and Fig.~\ref{fig:NSBH_kappa34}. The \ac{NSBH} results largely mirror the \ac{BNS} results: $\kappa_2(\bar X_f)$ remains smaller than $\kappa_2(\bar n_f)$ across the entire band, indicating that the variance of the cross-correlation estimator is still detector-noise-dominated. Nonetheless, we find $\sqrt{\kappa_2(\bar X_f)}/\sqrt{\kappa_2(\bar n_f)}\sim 30^{+13}_{-10}\%$ at $\sim\!14\,\mathrm{Hz}$, a level at which the contribution of $\kappa_2(\bar X_f)$ to $\kappa_2(\bar C_f)$ is no longer negligible; omitting it would lead to an underestimated analysis uncertainty or potential systematic biases. Furthermore, the non-Gaussianity remains negligible, with $|\gamma_1(\bar C_f)|\leqslant 2.2^{+1.8}_{-1.2}\times 10^{-3}$ and $\gamma_2(\bar C_f)\leqslant 1.3^{+1.0}_{-0.7}\times 10^{-4}$.

Overall, among the three CBC populations considered, only the \ac{BBH}-induced $\bar C_f$ exhibits significant non-Gaussianity in the \ac{XG} era, while the $\bar C_f$ from the \ac{BNS} and \ac{NSBH} backgrounds remains effectively Gaussian. In the near future, the non-Gaussianity of $\bar C_f$ can be ignored for all three types of astrophysical sources.


%% file: discussion.tex
\section{Discussion and conclusion}\label{sec:discussion_conclusion}
In this paper, we have characterized the statistical properties of the segment-averaged cross-correlation estimator $\bar C_f$ for an astrophysical \ac{SGWB} sourced by \ac{CBC} populations. By decomposing $\bar C_f$ into three physically distinct components — the mean intensity, geometrical shot noise, and polarization leakage — and computing the cumulants $\kappa_n(\bar C_f)$ up to fourth order, we have quantified for the first time the non-Gaussianity of $\bar C_f$ across a wide range of frequencies and detector sensitivities.

The principal finding is that the non-Gaussianity of $\bar C_f$ depends strongly on both the source population and the detector configuration. For all three \ac{CBC} populations, $\bar C_f$ remains effectively Gaussian under the upcoming LIGO O5 sensitivity, so existing Bayesian frameworks based on Gaussian likelihoods can be applied without modification. For a two-detector \ac{CE} network, however, the situation changes qualitatively: $\bar C_f$ from the \ac{BBH} background develops non-negligible non-Gaussianity at $f=20$ Hz $(|\gamma_1|=0.31^{+0.04}_{-0.03},\gamma_2=1.3^{+0.4}_{-0.3}$), while $\bar C_f$ from the \ac{BNS} and \ac{NSBH} populations remain effectively Gaussian. See Sec.~\ref{sec:sim} for caveats to this calculation for the \ac{BNS} background.

In the \ac{XG} era, more than $99\%$ of \ac{BBH} mergers throughout the observable Universe will be individually resolvable~\cite{Evans:2021gyd,ET:2025xjr}, so the \ac{SGWB} from \acp{BBH} may no longer be a primary scientific target. In this case, the \ac{BBH} contribution can simply be removed from the data via notching~\cite{Zhong:2022ylh,Zhong:2024dss,Zhong:2025gxk}. If, however, one wishes to retain all data and perform a joint Bayesian analysis of three \ac{CBC} populations, more sophisticated methods to model the large non-Gaussianity of $\bar C_f$ will need to be developed. 

A separate but equally important caveat concerns frequency correlations. Current analyses typically assume $\bar C_f$ to be statistically independent across frequency bins, so that the likelihood factorizes into a product of per-bin contributions. This assumption holds for a \ac{CGWB}, but for a \ac{SGWB} generated by \ac{CBC} events the power emitted in different frequency bins is strongly \textit{positively correlated}, owing to the chirp-like evolution of \ac{CBC} signals in the time-frequency domain. Neglecting these correlations is unlikely to bias the posterior means of the parameters of interest, but it leads to overly wide posteriors, since the analysis no longer makes optimal use of the available information.

We defer to future work a detailed investigation of the impact of $\Coffdiag$ on the statistical properties of $\bar C_f$, the development of more refined statistical methods for handling strong non-Gaussianity, and the inclusion of correlations between $\bar C_f$ and $\bar C_{f'}$ in the Bayesian inference likelihood.